# Combining Recurrent Neural Networks and Adversarial Training for Human Motion Modelling, Synthesis and Control

Zhiyong Wang*
Institute of Computing
Technology CAS
University of Chinese
Academy of Sciences

Jinxiang Chai†
Texas A&M University

Shihong Xia‡
Institute of Computing
Technology CAS

**ABSTRACT**

This paper introduces a new generative deep learning network for human motion synthesis and control. Our key idea is to combine recurrent neural networks (RNNs) and adversarial training for human motion modeling. We first describe an efficient method for training a RNNs model from prerecorded motion data. We implement recurrent neural networks with long short-term memory (LSTM) cells because they are capable of handling nonlinear dynamics and long term temporal dependencies present in human motions. Next, we train a refiner network using an adversarial loss, similar to Generative Adversarial Networks (GANs), such that the refined motion sequences are indistinguishable from real motion capture data using a discriminative network. We embed contact information into the generative deep learning model to further improve the performance of our generative model. The resulting model is appealing to motion synthesis and control because it is compact, contact-aware, and can generate an infinite number of naturally looking motions with infinite lengths. Our experiments show that motions generated by our deep learning model are always highly realistic and comparable to high-quality motion capture data. We demonstrate the power and effectiveness of our models by exploring a variety of applications, ranging from random motion synthesis, online/offline motion control, and motion filtering. We show the superiority of our generative model by comparison against baseline models.

**Index Terms:** Computing methodologies—Computer graphics—Animation; Computing methodologies—Computer graphics—Motion capture

## 1 INTRODUCTION

This paper focuses on constructing a generative model for human motion generation and control. Thus far, one of the most successful solutions to this problem is to build generative models from prerecorded motion data. Generative models are appealing for motion generation because they are often compact, have strong generalization ability to create motions that are not in prerecorded motion data, and can generate an infinite number of motion variations with a small number of hidden variables. Despite the progress made over the last decade, creating appropriate generative models for human motion generation remains challenging because it requires handling nonlinear dynamics and long-term temporal dependencies of human motions.

In this paper, we introduce an efficient generative model for human motion modeling, generation and control. Our key idea is to combine the power of recurrent neural networks (RNNs) and adversarial training for human motion generation, that generates synthetic motions from the generator using recurrent neural networks (RNNs) and refines the generated motion using an adversarial neural network which we call the "refiner network". **Fig** 2 gives an overview of our method: a motion sequence $X_{RNN}$ is generated with the generator $G$ and is refined using the refiner network $R$. To add realism, we train our refiner network using an adversarial loss, similar to Generative Adversarial Networks (GANs) [9] such that the refined motion sequences $X_{refine}$ are indistinguishable from real motion capture sequences $X_{real}$ using a discriminative network $D$. In addition, we embed contact information into the generative model to further improve the quality of the generated motions.

We construct the generator $G$ based on recurrent neural networks (RNNs). Recurrent neural networks (RNNs) are connectionist models that capture the dynamics of sequences via cycles in the network of nodes. Recurrent neural networks, however, have traditionally been difficult to train because it often contain millions of parameters and have vanishing gradient problems when they are applied to handle long term temporal dependencies. We address the challenge by using Long Short Term Memory network (LSTM) architecture [15], which recently has demonstrated impressive performance on tasks as varied as speech recognition [10, 12], language translation [26, 32, 1] , and image generation [13].

Our refiner network, similar to Generative Adversarial Networks (GANs) [9], is built upon the concept of game theory, where two models are used to solve a minimax game: a generator which samples synthetic data from the model, and a discriminator which classifies the data as real or synthetic. Thus far, GANs have achieved state-of-the-art results on a variety of generative tasks such as style transfer [8], 3D object generation [6], image super-resolution [22], image translation [18] and image generation [29, 31, 2].

Our final generative model is appealing to human motion generation. In our experiments, we show that motions generated by our model are always highly realistic. Our model is also compact because we don't need to preserve the original training data once the model is trained. Another nice property of the model is that the size of the learned model does not increase as more data is added into the training process. In addition, when we get new training data, we do not need to train the model from scratch. Instead, we can utilize the previous networks as a pre-train model and fine tune the model with the new training data.

We have demonstrated the power and effectiveness of our model by exploring a variety of applications, including motion generation, motion control and motion filtering. With our model, we can sample an infinite number of naturally looking motions with infinite lengths, create a desired animation with various forms of control input such as the direction and speed of a "running motion", and transform unlabeled noisy input motion into high-quality output motion. We show the superiority of our model by comparison against a baseline generative RNN model.

### 1.1 Contributions

Our work is made possible by a number of technical contributions:

- We present a new contact-aware deep learning model for data-

---
*e-mail: wangzhiyong@ict.ac.cn
†e-mail: jchai@cs.tamu.edu
‡e-mail: xsh@ict.ac.cn

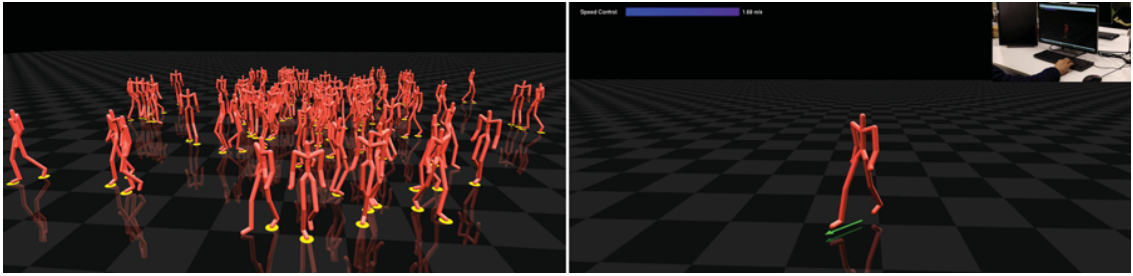

Figure 1: Human motion generation and control with our model. (left) random generation of high-quality human motions; (right) realtime synthesis and control of human motions.

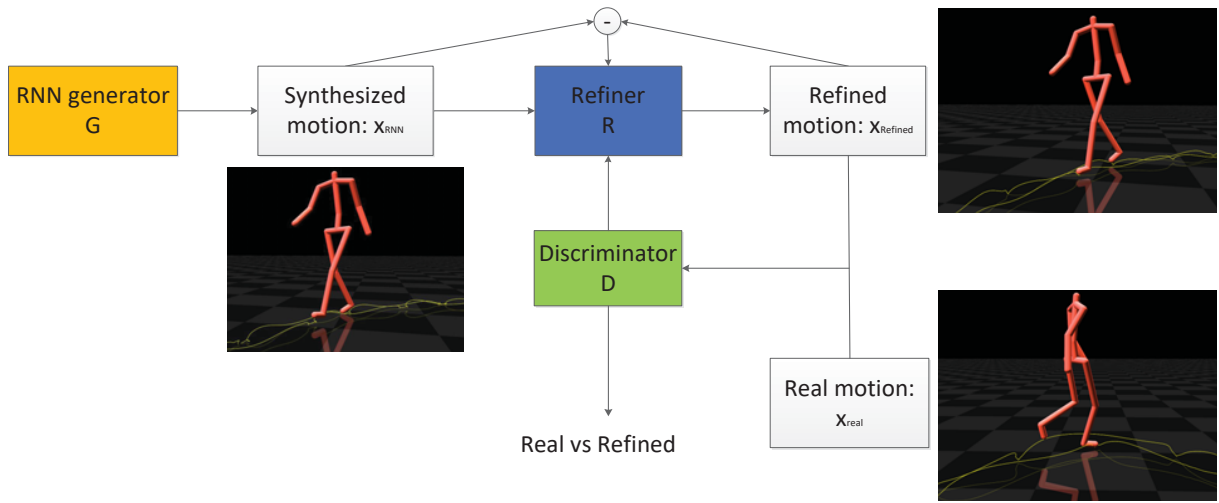

Figure 2: The Pipeline of our system. We first generate a motion sequence from the RNN generater, G, then refine the output of the generater with a refiner neural network, R, that minimizes the combination of a adversarial loss and a 'self-regularization' term. The adversarial loss 'fools' a discriminator network, D, that classifies the motion as real or refined. The self-regularization term minimizes the difference between the synthetic motion and the refined motion.

driven human motion modeling, which combines the power of recurrent neural networks and adversarial training.

- We train a adversarial refiner network to add realism to the motions generated by RNNs with LSTM cells using a combination of an adversarial loss and a self-regularization loss.
- We introduce methods for applying the trained deep learning model to motion synthesis, control and filtering.
- We make several key modifications to the deep learning framework to stabilize training and prevent the RNNs with LSTM and adversarial refiner network from producing artifacts.
- We will release the trained deep learning model, as well as training data sets, to the public.

## 2 BACKGROUND

Our approach constructs a generative deep learning model from a large set of prerecorded motion data and uses them to create realistic animation that satisfies various forms of input constraints. Therefore, we will focus our discussion on generative motion models and their application in human motion generation and control.

Our work builds upon a significant body of previous work on constructing generative statistical models for human motion analysis and synthesis. Generative statistical motion models are often represented as a set of mathematical functions, which describe human movement using a small number of hidden parameters and their associated probability distributions. Previous generative statistical models include Hidden Markov Models (HMMs) [3], variants of statistical dynamic models for modeling spatial-temporal variations within a temporal window [23, 5, 21, 28, 36], and concatenating statistical motion models into finite graphs of deformable motion models [27].

Most recent work on generative modeling has been focused on employing deep recurrent neural networks (RNNs) to model dynamic temporal behavior of human motions for motion prediction [7, 19, 25, 14, 4]. For example, Fragkiadaki and colleagues [7] proposed two architectures: LSTM-3LR (3 layers of Long Short-Term Memory cells) and ERD (Encoder-Recurrent-Decoder) to concatenate LSTM units to model the dynamics of human motions. Jain and colleagues [19] introduced structural RNNs (SRNNs) for human motion prediction and generation by combining high-level spatio-temporal graphs with sequence modeling success of RNNs. RNNs is appealing to human motion modeling because it can han-dle nonlinear dynamics and long-term temporal dependencies in human motions. However, as observed by other researchers [25, 14], current deep RNN based methods often have difficulty obtaining good performance for long term motion generation. They tend to fail when generating long sequences of motion as the errors in their prediction are fed back into the input and accumulate. As a result, their long-term results suffer from occasional unrealistic artifacts such as foot sliding and gradually converge to a static pose.

To address the challenge, we refine the motions generated by recurrent neural networks (RNN) using a "refiner network" with an adversarial loss, such that the refined motion sequences are indistinguishable from real motion capture data using a discriminative network. Adversarial training allows us to construct a generative motion model that can randomly generate an infinite number of high-quality motions with infinite length, a capability that has not been demonstrated in previous work. Our user studies show that the quality of motions generated by our model is comparable to high-quality motion capture data and is more realistic than those generated by RNNs. Our goal is also different from theirs because we aim to learn generative models for human motion synthesis and control rather than motion prediction for video-based human motion tracking. In our experiments, we show the user can create a desired animation with various forms of control input such as the direction and speed of a running motion.

Our work is relevant to recent efforts on character animation and control using deep learning methods [17, 16]. For instance, Holden et al. [17] trained a convolutional autoencoder on a large motion database and then learned a regression between high level parameters and the character motion using a feedforward convolutional neural network. In their more recent work, Holden et al. [16] constructed a simple three layer neural network to map the previous character pose and the current user control to the current pose, as well as the change in the phase, and applied the learned regression function for realtime motion control. Our model is significantly different from theirs because we combine RNNs and adversarial training to learn a dynamic temporal function that predicts the probability distributions of the current pose given the character poses and hidden variables in the past. Given the initial state of human characters, as well as the learned generative model, we can randomly generate an infinite number of high-quality motion sequences without any user input. Another difference is that we formulate the motion synthesis and control problem in a Maximum A Posteriori (MAP) framework rather than regression framework adopted in their work. The goal of our motion synthesis is also different because we aim to generate high-quality human motion from various forms of animation inputs, including constraints from offline motion design, online motion control and motion denoising while their methods are focused on for realtime motion control based on predefined control inputs.

Our idea of using adversarial training to improve the quality of synthesized motion from RNNs is motivated by the success of using an adversarial network to improve the realism of synthetic images using unlabelled real image data [31]. Specifically, Shrivastava and colleagues [31] proposed Simulated+Unsupervised (S+U) learning, where the task is to learn a model to improve the realism of a simulator's output using unlabeled real data, while preserving the annotation information from the simulator. They developed a method for S+U learning that uses an adversarial network similar to Generative Adversarial Networks (GANs), but with synthetic images as inputs instead of random vectors. We significantly extend their idea for S+U learning for image synthesis to human motion synthesis by improving the realism of human motion generated by RNNs with adversarial network using prerecorded human motion capture data.

## 3 OVERVIEW

Our goal herein is to learn generative networks from prerecorded human motion data and utilize them for generating naturally looking human motions that are consistent with various forms of input constraints. The whole system consists of two main components:

**Motion analysis.** We describe a method for learning the deep learning model from preprocessed motion data. To be specific, our first step is to learn a generative model based on RNNs with long short-term memory (LSTM) cells. Next, we train a refiner network using an adversarial loss such that the refined motion sequences are indistinguishable from real motion capture data using a discriminative network. We embed contact information into the GANs to further improve the performance of the GANs model.

**Motion synthesis.** We show how to apply the learned GANs to motion generation and control. We formulate the problem in a Maximum A Posteriori (MAP) framework. Given the initial state of human characters, as well as the learned generative model, we find the most likely human motion sequences that are consistent with control commands specified by the user and environmental constraints. We combine sampling-based methods with gradient-based optimization to find an optimal solution to the MAP problem. We discuss how to utilize the contact information embedded in the generative model to further improve the quality of output animation. We adopt a similar MAP framework to transform unlabeled noisy input motion into high-quality animation.

We describe details of each component in next sections.

## 4 MOTION ANALYSIS

Our goal is to develop a generative model that formulates the probabilistic distribution of the character state at the next frame $\mathbf{x}_{t+1}$ given the character state and hidden variables at the current frame, denoted as $\mathbf{x}_t$ and $\mathbf{h}_t$ respectively. Mathematically, we want to model the following probabilistic distribution:

$$p(\mathbf{x}_{t+1}|\mathbf{x}_t, \mathbf{h}_t) \qquad (1)$$

In the following, we explain how to model the probabilistic distribution using RNNs in Section 4.1 and how to train a refiner network using adversarial loss such that the refined motion sequences are indistinguishable from real motion capture data in Section 4.2. Section 4.3 discusses how to utilize the contact information embedded in the generative model to further improve the quality of output animation.

### 4.1 Generative RNNs Model

In this section, we first explain how to define the feature of the character state $\mathbf{x}_t$. Then, we give a brief introduction for RNNs and LSTM cells and explain how to apply RNNs with LSTM cells to generative motion modelling. In addition, we provide implementation details on how to stabilize training of the RNNs model.

#### 4.1.1 State Feature Representation

Each motion sequence contains trajectories for the absolute position and orientation of the root node(pelvis) as well as relative joint angles of 18 joints. These joints are head, thorax, and left and right clavicle, humerus, radius, hand, femur, tibia, foot and toe. Let $\mathbf{q}_t$ represent the joint angle pose of a human character at frame t, it can be written as:

$$\mathbf{q} = \begin{bmatrix} t_x\ t_y\ t_z\ r_x\ r_y\ r_z\ \theta_2\ \cdots\ \theta_d \end{bmatrix}^T \qquad (2)$$

where $t_x, t_y, t_z$ are the 3D position of root joint, $r_x, r_y, r_z$ are joint angles of the root joint, and $\theta_2, ..., \theta_d$ are joint angles of the other joints.

To define the feature of the character state, we choose to use relative rotation between current frame and previous frame for rotation around y axis. We name it $\triangle r_y$, which represents the relative global rotation of the character. For the same global rotation, our global rotation feature would be the same no matter which direction the character was facing to in the last frame. In other words, our feature is rotation invariant. Our translation feature in *x* and *z* axis are defined on the local coordinate of the previous frame. For the same global translation, our translation feature in *x* and *z* axis would be the same wherever the character is. This means our root translation feature for *x* and *z* axis are translation invariant.

The relationship between joint angle pose **q** and state feature **x** can be described as follows:

$$\mathbf{x} = \begin{bmatrix} \triangle t_x & \triangle t_z & \triangle r_y & t_y & r_x & r_z & \theta_2 & \cdots & \theta_d \end{bmatrix}^T \quad (3)$$

where the first three parameters $\triangle t_x, \triangle t_z, \triangle r_y$ are global features mentioned above, and $t_y, r_x, r_z$ are the other three components of the root joint.

Similarly, we can easily transform the state feature **x** back to the corresponding joint angle pose **q**. As our global rotation and translation is related to the previous frame, we use a homogeneous global transformation matrix to maintain the global status of the character and initialize it to the identity matrix. For every frame, we can get the local matrix of the frame from features and update the global transformation matrix by:

$$M_{t+1} = M_t M_{t+1,local}$$

$$M_{t+1,local} = \begin{bmatrix} & & & \Delta t_x \\ Rot_{3\times 3}(\Delta r_y) & & 0 \\ & & & \Delta t_z \\ 0 & 0 & 0 & 1 \end{bmatrix} \quad (4)$$

where the right column of $M_{t+1}$ contains the global root position on x-z plane for frame *t + 1*. The rotation of all the joints can be directly recovered from the motion feature.

### 4.1.2 Motion Modeling with RNNs

RNN is a deep neural network that has been widely used to model dynamic temporal behaviours. It has a parameter sharing over time by using the same group of parameter for every frame. In our application, it takes the hidden states and current feature as input and is trained to predict the probabilistic distribution of the feature at the next frame. The hidden states in RNN model carry the information about the history. This is the reason why RNN is suitable for handling long term temporal dependence. The derivative of RNN needs to be computed by Back Propagation Through Time (BPTT) [37] similar to normal back propagation methods except for its sequential structure. Like many other deep neural networks, RNN also suffers from the problem of vanishing gradient because the gradient flow must pass through an activation layer in each frame and thus the magnitude of gradient decreases quickly over time. This prevents the network from taking a relatively long history into account. LSTM cells [15] were introduced to address this challenge. LSTM cells divide hidden states into two parts: ***h*** is sensitive to short term memory, while ***C*** carries long term memory. Please refer to Appendix for more details of LSTM cells.

LSTM cells ensure that the long term memory $C_t$ is no longer influenced by activation functions. So the vanishing gradient problem is significantly reduced. As the LSTM model introduces a new variable $C_t$ that carries the long term memory, the formulation of the problem can be rewritten as following:

$$p(\mathbf{x}_{t+1}|\mathbf{x}_t, \mathbf{h}_t, \mathbf{C}_t) \quad (5)$$

The structure of our RNN motion model is shown in **Fig** 3. To predict the distribution of the next frame, our output should not be the state features itself. Similar to [11], we model the probabilistic distribution of the feature in the next frame using Gaussian Mixture Model (GMM). The distribution of GMM is written as follows:

$$p(\mathbf{x}_{t+1}) = \sum_{i=1}^{M} w_i N(\mathbf{x}_{t+1}|\mu_i, \sigma_i) \quad (6)$$

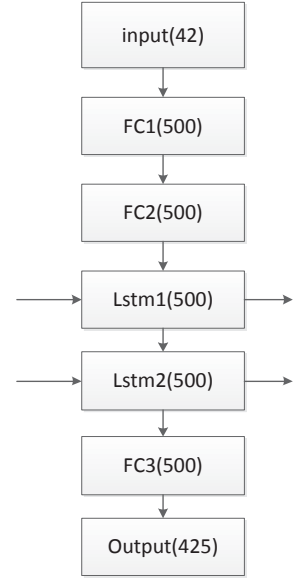

Figure 3: The structure of our RNN model. It has three fully connected layers (FC1,FC2,FC3) and two LSTM layers (LSTM1 and LSTM2). The numbers in the brackets are the width of the corresponding layer.

where the *i*-th vector component is characterized by normal distributions with weights $w_i$, means $\mu_i$ and standrard deviations $\sigma_i$. To simplify the problem, we assume that the covariance matrix of every Gaussian is a diagonal matrix. GMM model requires $\sum_{i=1}^{M} w_i = 1$ and $w_i > 0$, $\sigma_i > 0$. However, the output of our network can be in $(-\infty, +\infty)$. To bridge this gap, we define a transformation between the network output $\widehat{w_j}, \widehat{\mu}_{i,j}, \widehat{\sigma_{i,j}}$ and the GMM parameters $w_i, \mu_{i,j}, \sigma_{i,j}$ as follows:

$$w_i = \frac{e^{\widehat{w_i}}}{\sum_{j=1}^{M} e^{\widehat{w_j}}}, \mu_{i,j} = \widehat{\mu}_{i,j}, \sigma_{i,j} = e^{\widehat{\sigma_{i,j}}} \quad (7)$$

Here, $\sigma_{i,j}$ is the standard deviation of the j-th dimension in the i-th gaussian. By this transformation, we ensure the weights $w_i$ and standard deviations of Gaussian distribution $\sigma_{i,j}$ is positive, and $\sum_{i=1}^{M} w_i = 1$. M is set to 5 in our experiment.

The loss function for every frame is

$$\arg\min E = -\log p(\mathbf{x}_{t+1}|\mathbf{x}_t, \mathbf{h}_t, \mathbf{C}_t)) \quad (8)$$

During training, we compute $\frac{\partial E}{\partial p}$ from GMM transformation and back propagate the derivative to top layers to optimize the network parameters. Our network was trained with RMSProp [34]. In our implementation, every dimension of the derivative is clipped in range [-5, 5].

### 4.1.3 Model Training

As observed by other researchers [25, 14], RNN based generative models often have difficulty obtaining good performance for long term motion generation. They tend to fail when generating long sequences of motion as the errors in their prediction are fed back into the input and accumulate. As a result, their long-term results suffer from occasional unrealistic artifacts such as foot sliding and gradually converge to a static pose. In the following, we summarize our strategies on training RNNs:

**Adding noise into the training process.** Because the prediction of every frame often has small error, there is a risk that the error might accumulate over time and cause the model to crash at some point. Similar to [33, 11], we introduce independent identically distributed Gaussian noise to the state feature $\mathbf{x}_t$ during the training time in order to improve the robustness of our model prediction. By adding noise to the network input, the network to be learned becomes more robust to noise in the prediction process. As a result, the learned model becomes more likely to recover from small error in the prediction process. In our experiment, we set the mean and standard deviation of Gaussian noises to 0 and 0.05, respectively.

**Down sampling the training data.** We find that downsampling the training data from 120*fps* to 30*fps* can improve the robustness of the prediction model. We suspect that downsampling allows the RNN to handle longer term temporal dependence in human motions.

**Optimization method.** Optimization is critical to the performance of the learned model. We have found that RMSProp[34] performs better than simple SGD method adopted in [7].

**The size of the training data sets.** Training RNN model from scratch requires a lot of training data. Insufficient training data might result in poor convergence. We find that training an initial model with large diverse datasets, *(e.g., the CMU dataset)* and then refining the model parameters using a smaller set of high-quality motion data would reduce the convergence error and lead to better motion synthesis result.

The batch size is set to 20 and the time window size is set to 50. Our learning rate is initialized to be 0.001, it is multiplied by 0.95 after every epoch. We trained for 300 epochs. But we found that the loss converges after about 150 epochs, which is about 30000 iterations.

### 4.1.4 Motion Generation with RNNs

We now describe how to use the learned RNN to generate a motion instance. Given the character poses of the initial frames, $\mathbf{q}_0$ and $\mathbf{q}_1$, we first transform them into the feature space $\mathbf{x}_0$ and $\mathbf{x}_1$ as we described in Section 4.1.1. For every frame, given the current feature $\mathbf{x}_t$ and the current hidden states $\mathbf{h}_t$ and $\mathbf{C}_t$, we apply them as the input of the RNN model to obtain a Gaussian Mixture Model for the probabilistic distribution of the feature at the next frame(see **Fig** 3), the hidden states are updated in the same time. Next, we sample from the Gaussian Mixture Model to get an instance for the feature at the next frame $\mathbf{x}_{t+1}$. We repeat the process to generate a sequence of state features over time: $\mathbf{x}_0, \mathbf{x}_1, \mathbf{x}_2, ..., \mathbf{x}_n$. At last, we transform the state features back to the corresponding joint angle poses to create a motion sequence: $\mathbf{q}_0, \mathbf{q}_1, \mathbf{q}_2, \ldots, \mathbf{q}_n$.

### 4.2 Adversarial Refiner Training

Starting from the motion features $\mathbf{x}_{RNN}$ generated by generative RNN model $G$, our goal herein is to train a refiner network $R$ using an adversarial loss such that the refined motion data $R_\theta(\mathbf{x}_{RNN})$ are indistinguishable from real motion data $\mathbf{x}_{real}$ using a discriminative network $D$, where $\theta$ are the parameters of the refiner network.

To add realism to the motion generated by RNNs, we need to bridge the gap between the distributions of synthesized motion data and real motion data. An ideal refiner will make it impossible to classify a given motion sequence as real or refined with high confidence. This need motivates the use of an adversarial discriminator network, $D_\phi$, which is trained to classify motions as real vs refined, where $\phi$ are the parameters of the discriminator network. The adversarial loss used in training the refiner network is responsible for "fooling" the network into classifying the refined motions as real. Following the GAN approach, we model this as a two-player minimax game and update the refiner network $R_\theta$ and the discriminator network $D_\phi$ alternately.

#### 4.2.1 Generative Adversarial Networks

The adversarial framework learns two networks (a generative network and a discriminative network ) with competing losses. The generative model ($R$) (*i.e.*, the refiner network) transforms the motion generated by RNN ($\mathbf{x}_{RNN}$) into the refined motion $R_\theta(\mathbf{x}_{RNN})$ and tries to "fool" the discriminative model D. The loss function for the generative model is described as follows:

$$\arg\min_\theta -log(D_\phi(R_\theta(\mathbf{x}_{RNN}))) \qquad (9)$$

where $D_\phi(\mathbf{x})$ is the probability of $\mathbf{x}$ being classified by the discriminative network as real data. This loss function wants to "fool" the discriminative model, so that the refined motion $R_\theta(\mathbf{x}_{RNN})$ is indistinguishable from real motion data. We use the opposite version of the original GAN model [9] here to avoid early gradient vanish problem especially when the discriminator is too strong.

We define the loss function of the discriminative model as follows:

$$\arg\min_\phi -log(1-D_\phi(R_\theta(\mathbf{x}_{RNN}))) - log(D_\phi(\mathbf{x}_{real})) \qquad (10)$$

This loss function ensures the learned discriminative model is capable of distinguishing "real" or "refined" data.

In our implementation, we focus the refiner network on the features that are ignored by the RNN model. The most important features lost in the RNN model are the positions and velocities of end effectors. So we compute the velocities and positions of the end effectors from the input $\mathbf{x}$, denoted as $\mathbf{p}_{end}(\mathbf{x})$ and $\mathbf{v}_{end}(\mathbf{x})$, respectively, and include them into input for both the generative and discriminative model.

**Refiner network.** Our generative model has two fully connected layers and one LSTM layer (**Fig** 4). Mathematically, the refiner network learns a regression function $R_\theta$ that maps the input motion $\mathbf{x}$, as well as the positions and velocities of end effectors, to the refined motion $R_\theta(\mathbf{x})$:

$$\mathbf{R}_{input} = \begin{bmatrix} \mathbf{x} \\ \mathbf{p}_{end}(\mathbf{x}) \\ \mathbf{v}_{end}(\mathbf{x}) \end{bmatrix} \qquad (11)$$

$$R_\theta(\mathbf{x}) = (\mathbf{w}_{r2} \cdot lstm(relu(\mathbf{w}_{r1} \cdot \mathbf{R}_{input} + \mathbf{b}_{r1})) + \mathbf{b}_{r2}) + \mathbf{x}$$

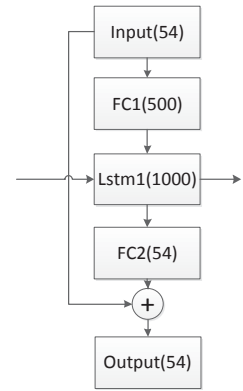

Figure 4: The structure of GAN generater network. It consists of two fully connected layers (FC1 and FC2) and one LSTM layer (Lstm1). The numbers in the brackets are the width of the corresponding layer.

Where the network parameters $\theta$ include weights of the LSTM layer (Lstm1) and weights and bias of two fully connected layers, including $\mathbf{w}_{r_1}, \mathbf{w}_{r_2}$ and $\mathbf{b}_{r_1}, \mathbf{b}_{r_2}$.

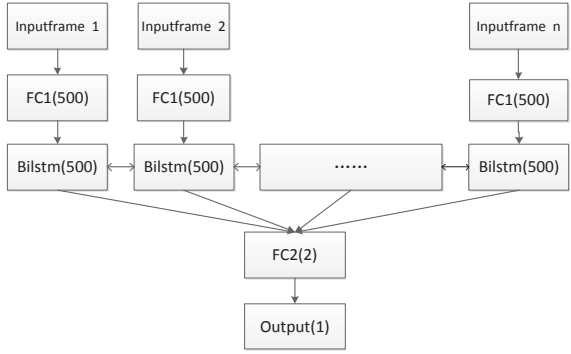

Figure 5: The structure of GAN discriminator network. It consists of one fully connected layers (FC1, FC2) and one bidirectional LSTM layer (Bilstm). The numbers in the brackets are the width of the corresponding layer.

**Discriminative model.** We apply a bidirectional LSTM [30] to model the discriminative model. The structure of the network is shown in **Fig** 5. Mathematically, it is defined as follows:

$$\mathbf{d}_{input} = \begin{bmatrix} \mathbf{p}_{end}(\mathbf{x}) \\ \mathbf{v}_{end}(\mathbf{x}) \end{bmatrix}$$
$$D_\phi(\mathbf{x}) = \mathbf{w}_{d2} \cdot bilstm^d(relu(\mathbf{w}_{d1} \cdot \mathbf{d}_{input} + \mathbf{b}_{d1})) + \mathbf{b}_{d2} \quad (12)$$

Where the network parameters $\phi$ include weights of the bidirectional LSTM layer and weights and bias of two fully connected layers, including $\mathbf{w}_{d_1}, \mathbf{w}_{d_2}$ and $\mathbf{b}_{d_1}, \mathbf{b}_{d_2}$.

**Motion regularization.** We add a regularization term to the generative loss function to ensure the difference between the input motion and refined motion is as small as possible. This leads to the following generative loss function:

$$\arg\min_\theta E = -log(D_\phi(R_\theta(\mathbf{x}_{RNN}))) + \lambda ||root(\mathbf{x}) - root(R_\theta(\mathbf{x}))||^2 \quad (13)$$

where $root(\mathbf{x})$ and $root(R_\theta(\mathbf{x}))$ are the root positions of the input motion $\mathbf{x}$ and the refined motion $R_\theta(\mathbf{x})$, respectively. And the weight $\lambda$ controls the importance of the regularization term. In our experiment, $\lambda$ is set to 20.

#### 4.2.2 Adversarial Training Details

GANs are very hard to train because of the competition between generative and discriminative model. It is fairly easy to break down when one of the two models is too strong. Our training strategies for GANs are summarized as follows:

**Training the generative model more.** We found that if the two networks are trained equally in a cycle, the discriminator often dominates the generator and leads the training to crash. In our practice, the generative model and discriminative model are each updated 75 times in the beginning. After that, we update the generative model five times and the discriminative model once at every step.

**Using a history of refined motions.** Another problem of adversarial training is that the discriminator network only focuses on the latest refined motions. The lack of memory may cause (i) false divergence of the adversarial training, and (ii) the refiner network reintroducing the artifacts that the discriminator has forgotten. To solve this problem, we update the discriminator using a history of refined motion rather than only the ones generated by the current network. To this end, similar to [31], we improve the stability of adversarial training by updating the discriminator using a history of refined motions, rather than only the ones in the current mini-batch. We slightly modify the algorithm to have a buffer of refined motions generated by previous networks. Let B be the buffer size, and b be the batch size. We generate B fake data in the very beginning. At each iteration of discriminator training, we sampling $\frac{b}{2}$ motions from the current refiner network, and sampling an additional $\frac{b}{2}$ motions from the buffer to update the parameters of the discriminator. After each training iteration, we randomly replace $\frac{b}{2}$ samples in the buffer with the newly generated refined motions. In practise, B is set to 320 and b is set to 32.

**Adjusting the training strategy when one of the models is too strong.** During training, we found that sometimes discriminative model is easy to be trained too strong that the generative model is nearly broken. To balance the training, we multiply the iteration times of the generative model by 2 when discriminator's softmax loss $< 0.01$. In contrast, we divide the iteration times of generative model by 2 when discriminator's softmax loss $> 1$. We found this to be useful to avoid crash. This approach is similar to [2].

Both the generative and discriminative models are trained by RMSProp [34]. We set the learning rate of the refiner to be 0.002 and the learning rate of the discriminator to be 0.005. The decay rate of RMSProp is set to 0.9.

#### 4.2.3 Motion Generation with Adversarial Training

After the refiner network is trained, we can combine it with the generative RNN model for motion synthesis. To achieve this goal, we first choose an initial state, we use it along with the RNN generative model (G) to randomly generate a motion sequence as described in section 4.1.4. Next, we compute the velocities and positions of the end effectors for every frame of the generated motion sequence. We augment the motion features with the velocities and positions of the end effectors and input them to the refiner network (R) to obtain the refined motion features. In the final step, the refined motion features are transformed back to the corresponding joint angle poses to form the output animation.

### 4.3 Contact-aware Motion Model

We now describe how to embed the contact information into our generative model to further improve the quality of the generated motion. We encode contact information in each frame using a binary feature vector. Each bit represents a particular type of contact events, e.g., left toe plants on the ground. In our application, we encode the contact information into a 2x1 binary vector $\mathbf{c} = [c_l, c_r]^T$. The first bit represents if the left foot of the character is on the ground and the second bit represents if the right foot is on the ground.

We augment the motion feature with contact information for RNN modelling. In addition, we augment both the network input and the network output with contact vector in adversarial training. Contact awareness further improve the quality of generated motion. Another advantage of contact-awareness is to automatically label every frame of the generated motion with contact information. This allows us to enforce environmental contact constraints in a motion generalization process, thereby eliminating noticeable visual artifacts such as foot sliding and ground penetration in output animation. **Fig** 6 shows a comparison before and after the refinement.

## 5 MOTION SYNTHESIS AND CONTROL

In this section, we demonstrate the power of our GAN motion model for various applications, including random motion generation, offline motion design, online motion control, and motion denoising.

### 5.1 Random Motion Generation

To do motion control accurately, a motion model must be generative model which can generate varies motions. We first show our motion's ability of generating different motions. As mentioned in 4.1.4 and 4.2.3, our model can sample varies motions from any initial frame. Similar to what we did in training, we allow the input of network to have small noise. Experiment show it would expand the generative ability of the model. This means we can have network noise $\mathbf{d}_1, \mathbf{d}_2,...,\mathbf{d}_n$ as another group of variables in addition to motion features. The generated motion is foot-contact aware because of the contact-awareness of our model. **Fig** 7 shows an example.

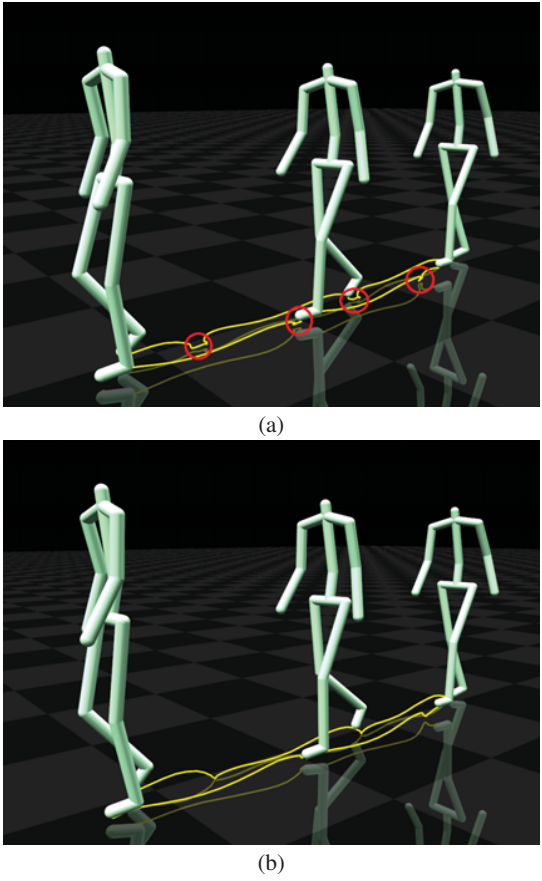

(a)

(b)

Figure 6: A comparison between (a) motion generated by RNN and (b) motion generated by refinement network. GAN model eliminates the foot sliding problem and get a more robust performance.

### 5.2 Offline Motion Design

Our generative model is also well suited for generating naturally looking human motion consistent with user-defined input. We formulate the motion synthesis and control problem in a MAP estimation framework. Given user-defined constraints $\mathbf{c}$ and environmental constraint $\mathbf{e}$ for the sequence, as well as initial motion state $\mathbf{s}_0$, we want to find a sequence of motion vectors $\mathbf{s}=\{\mathbf{s}_i, i=1,2,...T\}$ which is most likely to appear. That means we want to maximize the probability of $P(\mathbf{s},\mathbf{d}|\mathbf{s}_0,\mathbf{c},\mathbf{e})$. According to Bayes rule, we have:

$$\begin{aligned}&\arg\max_{\mathbf{s},\mathbf{d}} P(\mathbf{s},\mathbf{d}|\mathbf{s_0},\mathbf{c},\mathbf{e})\\&=\arg\max_{\mathbf{s},\mathbf{d}} \frac{P(\mathbf{s},\mathbf{d}|\mathbf{s_0})P(\mathbf{c},\mathbf{e}|\mathbf{s}_0,\mathbf{s},\mathbf{d})}{P(\mathbf{c},\mathbf{e}|\mathbf{s}_0)}\\&\propto \arg\max_{\mathbf{s},\mathbf{d}} P(\mathbf{s},\mathbf{d}|\mathbf{s}_0)P(\mathbf{c}|\mathbf{s}_0,\mathbf{s},\mathbf{d})P(\mathbf{e}|\mathbf{s}_0,\mathbf{s},\mathbf{d})\end{aligned} \quad (14)$$

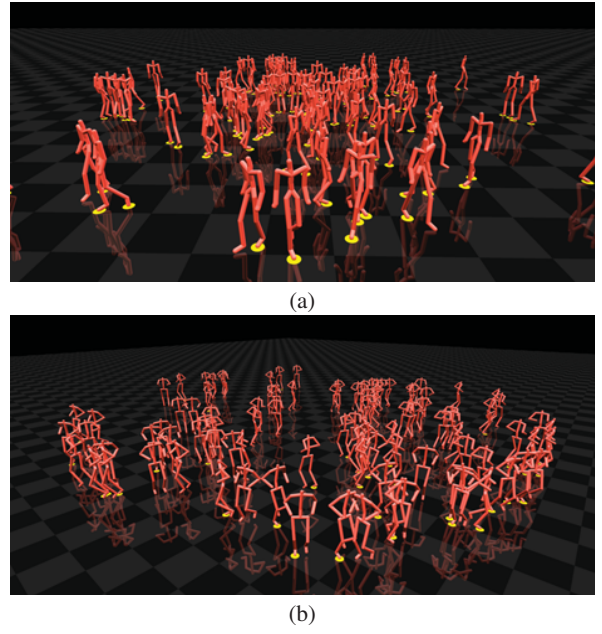

(a)

(b)

Figure 7: The random motion generation results of: (a) walking motion and (b) running motion. The yellow circles show foot contact label.

The first term is the prior term, it can be written as:

$$p(\mathbf{s},\mathbf{d}|\mathbf{s}_0) = P(\mathbf{d}_1,\mathbf{d}_2,...,\mathbf{d}_T)\prod_{i=1}^{T}P(\mathbf{s}_i|\mathbf{s}_0,\mathbf{s}_1,...,\mathbf{s}_{i-1},\mathbf{d}_0,...,\mathbf{d}_{i-1}) \quad (15)$$

The first part of the prior term is a Gaussian noise, it obeys Gaussian distribution, the prior for noise is:

$$P(\mathbf{d}_1,\mathbf{d}_2,...,\mathbf{d}_n) = \prod_{i=2}^{n} e^{-\frac{\mathbf{d}_i^2}{2\sigma_{noise}^2}} \quad (16)$$

the $\sigma_{noise}$ here is set to 0.05 by experiment. The second part of the prior is the probability from the RNN-based GAN models.

**Control term.** The second term of equation (14) is the control term. Our model enables the user to accurately control a character at the kinematic level. Low-level kinematic control is important because it allows the user to accurately control motion variations of particular actions. Our motion control framework is very flexible and supports any kinematic control inputs. The current system allows the user to control an animation by selecting a point (*e.g., root*) on the character and specifying a path for the selected point to follow. The user could also direct the character by defining the high-level control knobs such as turning angles, step sizes, and locomotion speeds. More specifically, we allow the user to control the root path of an animated character. For root joint projection on the ground for every frame $root(\mathbf{s}_i)$, we first find the nearest points $\mathbf{c}_{near,i}$ on the curve. We assume there is a Gaussian noise with a standard deviation of $\sigma_{fit}$ for the user's control inputs $\mathbf{c}$. Then we can define the likelihood of fitting as follow:

$$p_{curve} \propto \prod_{i=1}^{T} e^{-\frac{\|root(\mathbf{s}_i) - \mathbf{c}_{near,i}\|^2}{\sigma_{fit}^2}} \quad (17)$$

The standard deviation of fitting term $\sigma_{fit}$ indicates the preference of the user's fitting accuracy. The smaller it is, the controller would pay more attention to fitting accuracy. When the given curve is not achievable and $\sigma_{fit}^2$ is too small, the synthesized motion would be strange. In our experiment, $\sigma_{fit}^2$ is set to 0.5.

**Contact awareness term.** The last term is the contact awareness term. Due to the contact awareness of our model, our generated motion is automatically annotated with contact information. We first retrieve contact information from the network output. Then for each frame, if there is a contact between the character and the environment, we measure two distances: the first one is the distance between the synthesized contact point on the character of current frame and previous frame; the second one is the point plane distance between the synthesized contact point on the character and the corresponding contact plane. We assume Gaussian distributions with a standard deviation $\sigma_{con}$ for adjacent of the contact-awareness term, and $\sigma_{con,y}$ for point-plane constraint of the contact-awareness term. Then the contact awareness term can be written as:

$$P_{contact} = e^{-\frac{\|f(s_{i-1})_{foot} - f(s_i)_{foot}\|^2}{\sigma_{con}^2}} \\ \cdot e^{-\frac{\|\mathbf{n} \cdot (f(s_{i-1})_{foot} - p_{plane})\|^2}{\sigma_{con,y}^2}} \quad (18)$$

Because the original probability is hard to evaluate, we transform to its -log form, the overall loss function is

$$\min_{\mathbf{s}_1,\mathbf{s}_2,...,\mathbf{s}_n,\mathbf{d}_1,\mathbf{d}_2,...,\mathbf{d}_n} E = -log(P_{prior} \cdot P_{noise} \cdot P_{curve} \cdot P_{contact}) \quad (19)$$

This loss function is non-convex because of the complexity of RNN. To solve this optimization problem, we combine the power of sampling and gradient based optimization to get good result in acceptable time. Given a start frame and a user defined curve, our experiment show that design the motion sequences frame by frame won't get good result. It's because in RNN model, the performance in every frame is highly related to previous frames. So we use spatial temporal optimization to make the generated motion sequence smooth. We achieve the synthesis by sliding window method. Each window has 34 frames with an overlap of 17 frames between the adjacent windows.

For each window, we first sample a certain number of motions each of which has a length of window size, then select a best initialization among them, then we do gradient based optimization to get more precise result. We use sampling based method to get a good initialization in the complex solution space. In detail, We take advantage of particle swarm optimization [20], which can get better result in the same time compare to random sampling. Particle swarm optimization can lead to good result when the optimization has converged, but it takes too long time. To accelerate the progress, we switch to gradient based optimization after the loss function is rather small. As our motion feature is based on previous global motion states(we use $\triangle r_y, \triangle t_x, \triangle t_z$ as part of features), our jacobian matrix for constraints term and foot contact term is much complex than previous approaches, we analytically evaluate it by *BPTT*. Because of the RNN prior term, our problem is not a least squares form. We use LBFGS [24] method to complete the optimization.

### 5.3 Online Motion Control

Besides offline motion design, our model is also suited for online control. We allow the user to control the speed and direction of the character(see **Fig** 9). Similar to offline case, we also model the problem in a MAP framework.

$$\arg\max_{\mathbf{s},\mathbf{d}} P(\mathbf{s},\mathbf{d}|\mathbf{s}_0)P(\mathbf{c}|\mathbf{s}_0,\mathbf{s},\mathbf{d})P(e|\mathbf{s}_0,\mathbf{s},\mathbf{d}) \quad (20)$$

We still have the RNN prior term, the fitting accuracy term and the contact awareness term. One difference is in the control term, we need to deal with direction and speed of the character. We also do the optimization in a length n batch each time. As a result, the control response would has a latency. As the control term is not the same, we formulate new loss function for the control terms. Similarly, our realtime motion control system allows for accurate and precise motion control at the kinematic level. Our current implementation allows the user to control the speed and direction of locomotion such as walking and running. The online constraints includes speed control and direction control, which means:

$$P(\mathbf{c}|\mathbf{s}_0,\mathbf{s},\mathbf{d}) = P(\mathbf{c}_{speed}|\mathbf{s}_0,\mathbf{s},\mathbf{d})P(\mathbf{c}_{direction}|\mathbf{s}_0,\mathbf{s},\mathbf{d}) \quad (21)$$

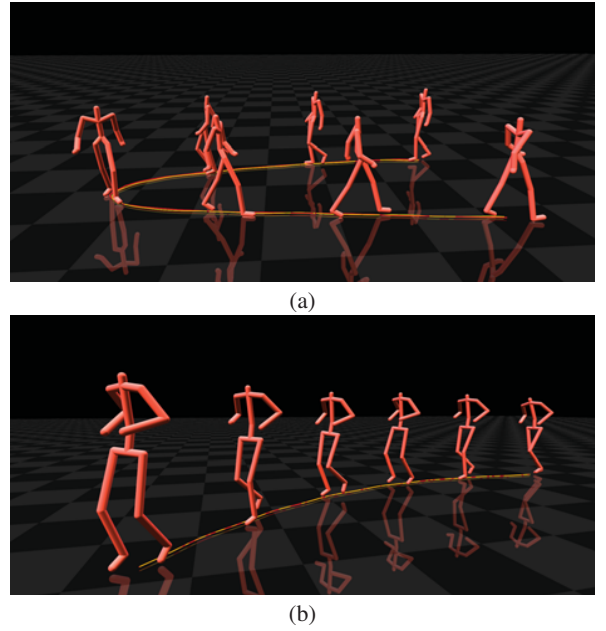

(a)

(b)

Figure 8: The result of offline motion control. The red line on the ground is the user input constraint, the yellow line is the root trajectory for the synthesized motion. We draw a few keyframe of the synthesized motion here. (a) is a walking result, (b) is a running result. Our results are best seen in video form.

We deal with speed and orientation online control in following.

**Speed control.** We allow the user to give a control speed for the character. As the speed of the character naturally changes by time, we constrain the character's average speed over the optimization batch as close as possible to the given speed control. Assuming the control result is a Gaussian distribution around the control input with standard deviation $\sigma_{speed}$, we have:

$$P(\mathbf{c}_{speed}|\mathbf{s}_0,\mathbf{s},\mathbf{d}) \propto e^{\frac{\|\frac{1}{T}\sum_{i=1}^{T} speed_i - \mathbf{c}_{speed}\|^2}{\sigma_{speed}^2}} \quad (22)$$

**Direction control.** The user can also control the moving direction of the character. We define the facing direction of the character as the angle around y axis of the root joint. We measure the difference between the direction of the last frame and the control input. We assume a gaussian distribution around the control input:

$$P(\mathbf{c}_{direction}|\mathbf{s}_0,\mathbf{s},\mathbf{d}) \propto e^{\|direct_T - \mathbf{c}_{direction}\|^2} \quad (23)$$

Here $direct_T$ is the direction of the last frame in the batch, $\mathbf{c}_{direction}$ is the direction control input. The optimization problem is also solved by gradient based optimization. Using GPU acceleration by a GTX970 graphics card, we have an average frame rate of 30fps.

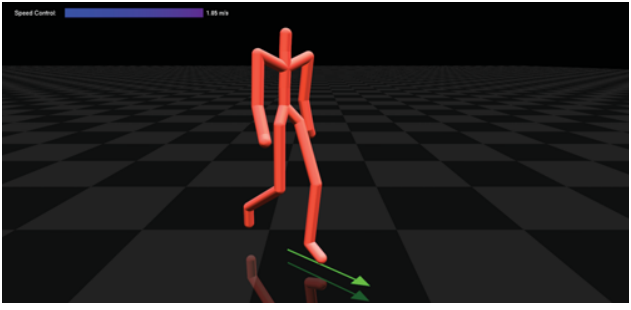

Figure 9: The online motion control. We have the speed and direction control here. The control speed is show on the left-top of the screen, the direction is shown by the arrow on the ground. Our results are best seen in video form.

### 5.4 Motion Denoising

Motion denoising takes a noisy and unlabeled motion sequence as an input and generates a naturally looking output motion that is consistent to the input motion (see **Fig** 10). We formulate the problem as a MAP problem:

$$\arg\max_{\mathbf{s},\mathbf{d}} P(\mathbf{s},\mathbf{d}|\mathbf{s}_0)P(\mathbf{c}|\mathbf{s}_0,\mathbf{s},\mathbf{d})P(\mathbf{e}|\mathbf{s}_0,\mathbf{s},\mathbf{d}) \quad (24)$$

The first prior term and the third environment contact constraints term is the same as that in section 5.2. We model the second constraint term in the following.

We assume the noise consists of two parts: the noise of the root position and the noise of the joint angles. We model both of the two type of noises in the original motion as a Gaussian noise. So we have:

$$P(\mathbf{c}|\mathbf{s}_0,\mathbf{s},\mathbf{d}) \propto e^{\frac{(\mathbf{c}_{root}-root(\mathbf{s}_i))^2}{2\sigma_{root}^2}} \cdot e^{\frac{\sum_{j=1}^{n}||angle_j(\mathbf{c})-angle_j(\mathbf{s}_i)||^2}{2\sigma_{angle}^2}} \quad (25)$$

Here $\mathbf{c}_{root}$ and $\mathbf{c}_{angle}$ and the root position and joint angles of the original motion, $root(\mathbf{s}_i)$ and $angle(\mathbf{s}_i)$ are the root position and joint angles of the filtered motion. This optimization problem is also solved by LBFGS [24].

**Initial state estimation.** There is a disadvantage for traditional RNN based motion model in the first frame. When we do motion synthesis, the RNN output would have a hopping between the first frame and the second frame. For most applications we can simply throw the first frame to escape from the problem, however, certain scenes (e.g. motion filter) request us to solve the problem. Traditional RNN based sequence synthesis method start the synthesis with zeros state [10] [7]. But in training, the input states for most frames is nonzero. The network is trained to fit the next frame with previous frame feature and appropriate hidden state. So when the hidden state of the network is set to zero in the first frame, the network is not trained enough to deal with this situation and may cause wrong result for synthesis. To solve the problem, we estimate what the hidden states should be for the first frame by RNN probability. We estimate the initial hidden state by gradient based optimization, the loss function is:

$$\min_{h_1} E = -log \sum_{i=1}^{M} w_i P(\mathbf{s}_2|\mathbf{s}_1, \mathbf{h}_1) \quad (26)$$

Here $h_1$ is the hidden states for the first frame. We compute the gradient $\frac{\partial E}{\partial h}$ by back propagation and use gradient descent method to do the optimization. Given appropriate initial states, the generated motion would be smooth in the first frame.

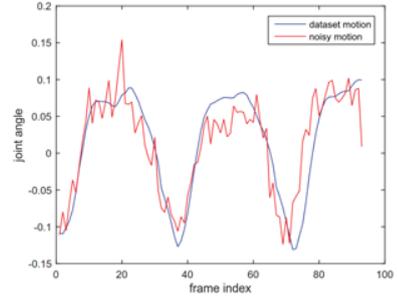

(a)

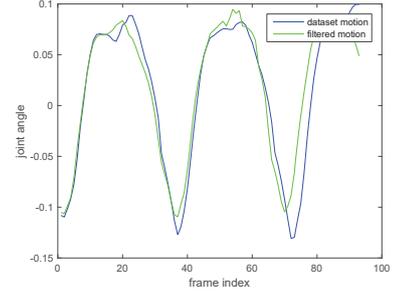

(b)

Figure 11: A comparison of one joint angle for the motion before and after the filter and a similar motion in the training dataset. The filtered motion is more smooth and similar to the dataset motion

## 6 RESULTS

We have demonstrated the power and effectiveness of our model on motion generation, motion control and motion filtering. To the best of our knowledge, this is the first generative deep learning model that is capable of generating an infinite number of high-quality motion sequences with infinite lengths. The user studies show motions generated by our model are comparable to motion capture data obtained by ten vicon cameras. In addition, we have demonstrated the superiority of our model over a baseline RNN model. Our results are best seen in video form.

We captured walking and running data of a single actor for 525 motion sequences. Those motion sequences varies in speed, step length, and turning angle. Besides, we use CMU dataset to do pre-train. Our model is relatively small because we do not need to preserve the original motion sequence once the model is trained. Our generative model (41Mb) is much smaller than the size of the original training data sets (133Mb). The original dataset contain 419488 frames (58.26 minutes) walking and running data varies in speed, stepsize, and turning angles. What is more, as the training data increases, the size of the model would not increase. When we get new training data, we don't need to train the model from stretch again. We can utilize the previous network as a pre-train model and fine tune it by all the training data.

**Random Motion Generation.** This experiment demonstrates our model can generate an infinite number of high-quality animation sequences with infinite length. We pick a certain starting frame, then genenrate 100 motions as the method we described in 4.2.3. Results show that the generated motion varies in step size, speed, and turning angles. Results can been seen as **Fig** 1.

**Offline Motion Design.** The user can generate a desired animation that consistent with control input. This experiment shows that our model can generate a desired animation by specifying the projection of the root trajectories on the ground. In the accompany-ing video, we show the process of creating desired, natural-looking animations with both hand drawn curves and pre-defined curves. Results can been seen in **Fig** 8.

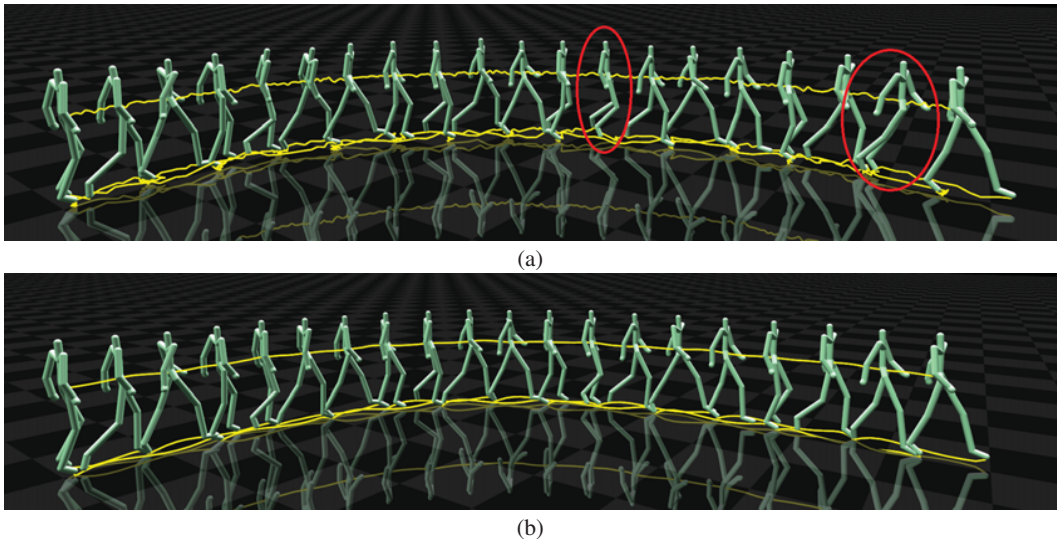

Figure 10: A comparison of motion before and after the filter. The trajectory of root joint and feet joints are shown in yellow. (a) the original noisy motion. (b) the motion after RNN-based GAN filter. The joint trajectory of the motion before filtering is jerky, outlier poses are circled by red. This result is best seen in video form.

**Online Motion Control.** The online motion control system offers precise, realtime control over human characters with the speed and direction of walking and running. The accompanying video shows that the character can make a sharp turn, e.g., -180-degree or +180-degree, or speed up and down with little latency. A result can be seen in **Fig** 9.

**Motion Denoising.** The accompanying video shows a side-by-side comparison between input noisy motion and output motion after denoising for both walking and running. In both cases, the input motion appears very jerky and contains significant foot sliding artifacts and occasional outlier poses. It is hard to judge when and where the foot contact occurs in the input motion. Our motion denoising algorithm automatically identifies the foot contact frames and output high-quality motion that closely matches the input motion. One joint angle of the motion before and after the filter can been seen in **Fig** 11. The keyframes of the motion before and after motion filter can be seen in **Fig** 10.

**The advantage of combining RNNs and adversarial training.** The accompanying video shows a side-by-side comparison of generated motions from RNNs with and without adversarial training. For both walking and running cases, the results from RNNs are occasionally jerky and contain frequent foot sliding artifacts while the combined model produces highly realistic motion without any noticeable visual artifact. This can also be seen in **Fig** 6.

**Motion quality.** We evaluate the quality of synthesized motions via user studies. To be specific, we compare the quality of our synthesized motions against high-quality motion capture data ("Mocap") and those generated from RNNs ("RNN"). We implement the "RNN" synthesis method based on the algorithm described in Section 4.1.4. "Mocap" represents high-quality motion capture training data captured by ten Vicon[35] cameras. "GAN1" and "GAN2" represent motions generated by a combination of our RNNs and refiner network with and without automatic foot contact handling, respectively. Note that our contact-aware deep learning model can automatically label every frame of the generated motion with contact information. This information allows us to automatically enforce environmental contact constraints in output motion.

We have evaluated the quality of motions on 25 users, including males and females. Most of the users have no previous experience with 3D animation. For both running and walking, we randomly generated seven animation sequences from each algorithm along with seven animation sequences randomly chosen from the mocap database. We render animations on a stick figure similar to **Fig** 9. Then we randomly organized all the animation clips. We asked the users to watch each video and provide a score of how realistic the motion is, ranged from 1 ("least realistic") to 5 ("most realistic"). We report mean scores and standard deviations for motions generated by each method, see **Fig** 12. The user studies show that the motions generated by our method ("GAN2") are highly realistic and comparable to those from the mocap database ("Mocap"). The evaluation also shows the advantage of combining RNNs and adversarial training for human motion synthesis as both "GAN1" and "GAN2" produce more realistic results than "RNN". In addition, the score difference between "GAN1" and "GAN2" shows the advantage of embedding contact information into deep learning model for human motion modeling and synthesis.

## 7 CONCLUSION AND FUTURE WORK

We have introduced a generative deep learning model for human motion synthesis and control. Our model is appealing for human motion synthesis because it is generative and can generate an infinite number of high-fidelity motion sequences to match various forms of user-defined constraints. We have demonstrated the power and effectiveness of our models by exploring a variety of applications, ranging from random motion synthesis, offline and realtime motion control, and motion denoising.

We have shown that motions generated by our model are always highly realistic. One reason is that our generative model can handle both nonlinear dynamics and long term temporal dependences of human motions. Our model is compact because we do not need to preserve the original training data once the model is trained. Another nice property of the model is that the size of our model does not increase as more data is added into the training process. Our model is also contact aware and embedded with contact information, thereby removing unpleasant visual artifacts often present in the motion

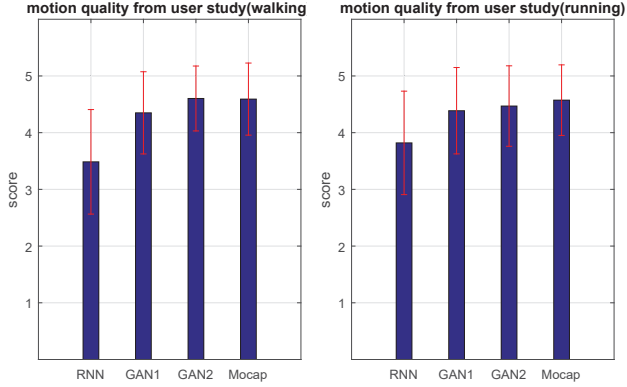

Figure 12: Comparisons of motion quality. We asked the users to give a score (1-5) of how realistic synthesized motions are. This graph shows the results of these scores, including means and standard deviations for generated motions via our methods with and without contact handling ("GAN1" and "GAN2"), a baseline RNNs method ("RNN"), and high-quality motion capture data from ten vicon cameras ("Mocap").

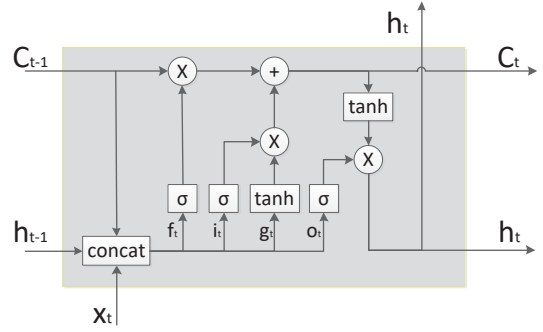

Figure 13: The structure of LSTM.

generalization process.

We formulate the motion control problem in a maximum a posteriori (MAP) framework. The MAP framework provides a principled way to balance the tradeoff between user constraints and motion priors. In our experiments, the input constraints from motion control might be noisy and could even be unnatural. When this happens, the system prefers to generate a "natural looking" motion that "'best" matches the input constraints rather than generating the "best" possible motion that "exactly" matches the user constraints.

We have tested the model on walking and running data set. In the future, we plan to test the model on aperiodic motions such as jumping and dancing. We also plan to test the model on a heterogeneous motion database such as walking, running, jumping and their transitions. Our system achieves realtime (30fps) on a GTX970 card, however, it is still not fast enough for mobile applications. Therefore, one direction for future work is to speed up the system via network structure simplification and model compression.

We show that our generative deep learning motion model can be applied for motion generation, motion control and filtering. We believe the model could also be leveraged for many other applications in human motion analysis and processing, such as video-based motion tracking, motion recognition, and motion completion. One of the immediate directions for future work is, therefore, to investigate the applications of the models to human motion analysis and processing.

## APPENDIX

## A THE STRUCTURE OF LSTM

LSTM cells divide hidden states into two parts: $h$ is sensitive to short term memory, while $C$ carries long term memory. Mathematically, it is:

$$(i, f, o, g)_t = W_{i,f,o,g} \cdot \begin{bmatrix} C_{t-1} \\ h_{t-1} \\ x_t \end{bmatrix} + b_{i,f,o,g}$$

$$\widetilde{i}_t = \sigma(i_t)$$
$$\widetilde{f}_t = \sigma(f_t) \qquad (27)$$
$$\widetilde{o}_t = \sigma(o_t)$$
$$\widetilde{g}_t = tanh(g_t)$$
$$C_t = \widetilde{f}_t \cdot C_{t-1} + \widetilde{i}_t \cdot \widetilde{g}_t$$
$$h_t = \widetilde{o}_t \cdot tanh(C_t)$$

Here $\sigma$ is the sigmoid function $\sigma(x) = \frac{1}{1+e^{-x}}$, and $tanh(x) = \frac{1-e^{-x}}{1+e^{-x}}$. $\widetilde{f}_t$ is called the forget gate. When $\widetilde{f}_t$ is rather small, all the history memory of the LSTM cell will loss. It is useful when we don't want the memory. $\widetilde{i}_t$ is called the input gate. New information get into the long term memory $\mathbf{C}$ by this gate. $o_t$ is called the output gate. It influences the output of the LSTM cell.